\documentclass[%
aip,
sd,%
amsmath,amssymb,
reprint,%
]{revtex4-1}

\usepackage{graphicx}
\usepackage{dcolumn}
\usepackage{bm}

\usepackage{amssymb}
\usepackage{color}

\newcommand{\umux}{$\mu$mux} 

\begin{document}
\title{Microwave SQUID Multiplexer Demonstration for Cosmic Microwave Background Imagers}
\author{B.~Dober}
\email{bradley.dober@nist.gov \\ Contribution of the National Institute of Standards and Technology; not subject to copyright in the United States}
\affiliation{National Institute of Standards and Technology, Boulder, CO 80305, USA}
\author{D.T.~Becker}
\affiliation{University of Colorado Boulder, Boulder, CO 80309, USA}
\author{D.A.~Bennett}
\affiliation{National Institute of Standards and Technology, Boulder, CO 80305, USA}
\author{S.A.~Bryan}
\affiliation{School of Earth and Space Exploration, Arizona State University, Tempe, AZ 85281, USA}
\author{S.M.~Duff}
\affiliation{National Institute of Standards and Technology, Boulder, CO 80305, USA}
\author{J.D.~Gard}
\affiliation{University of Colorado Boulder, Boulder, CO 80309, USA}
\author{J.P.~Hays-Wehle}
\affiliation{National Institute of Standards and Technology, Boulder, CO 80305, USA}
\affiliation{NASA Goddard Space Flight Center, Greenbelt, MD 20771, USA}
\author{G.C.~Hilton}
\affiliation{National Institute of Standards and Technology, Boulder, CO 80305, USA}
\author{J.~Hubmayr}
\affiliation{National Institute of Standards and Technology, Boulder, CO 80305, USA}
\author{J.A.B.~Mates}
\affiliation{University of Colorado Boulder, Boulder, CO 80309, USA}
\author{C.D~Reintsema}
\affiliation{National Institute of Standards and Technology, Boulder, CO 80305, USA}
\author{L.R.~Vale}
\affiliation{National Institute of Standards and Technology, Boulder, CO 80305, USA}
\author{J.N.~Ullom}
\affiliation{National Institute of Standards and Technology, Boulder, CO 80305, USA}
\affiliation{University of Colorado Boulder, Boulder, CO 80309, USA}

\begin{abstract}
Key performance characteristics are demonstrated for the microwave SQUID multiplexer (\umux) coupled to transition edge sensor (TES) bolometers that have been optimized for cosmic microwave background (CMB) observations.  
In a 64-channel demonstration, we show that the \umux\ produces a white, input referred current noise level of 29~pA$/\sqrt{\mathrm{Hz}}$ at -77~dB microwave probe tone power, which is well below expected fundamental detector and photon noise sources for a ground-based CMB-optimized bolometer.  
Operated with negligible photon loading, we measure 98~pA$/\sqrt{\mathrm{Hz}}$ in the TES-coupled channels biased at 65\% of the sensor normal resistance.  
This noise level is consistent with that predicted from bolometer thermal fluctuation (i.e. phonon) noise.
Furthermore, the power spectral density is white over a range of frequencies down to $\sim$~100~mHz, which enables CMB mapping on large angular scales that constrain the physics of inflation. 
Additionally, we report cross-talk measurements that indicate a level below 0.3\%, which is less than the level of cross-talk from multiplexed readout systems in deployed CMB imagers. 
These measurements demonstrate the \umux\ as a viable readout technique for future CMB imaging instruments.

\end{abstract}

\maketitle


To enable precise probes of the inflationary epoch and other scientific goals, the next generation of cosmic microwave background (CMB) imagers will utilize ever larger arrays of cryogenic sensors to achieve the requisite sensitivity.
Detector counts approaching $10^6$ are the baseline for ``CMB Stage IV,'' an experiment with multiple ground-based telescopes planned for the 2020s \cite{s4-science,s4-instrument}.
The large number of sensors presents a challenge for cryogenic readout. 
Modern cryogenic sensor arrays require multiplexed readout to reduce wiring complexity, cost, and thermal loads.  
Multiplexing combines signals from many detector channels into a smaller number of wires and readout amplifiers.


Currently deployed CMB imagers that use transition-edge-sensor (TES) bolometers employ one of two well-established multiplexing techniques which are based on signal amplification by use of superconducting quantum interference devices (SQUIDs): time-division\cite{TDMmux} and megahertz frequency-division\cite{FDMmux} multiplexing.  
Both architectures have been used to read out cameras with $> 10^4$ detectors \cite{scuba2,spt3g}; however the maximum currently implemented number of sensors per amplifier chain (i.e., the multiplexing factor) in a fielded instrument is $\sim$\,70,\cite{spt3g,henderson2016} and 160 has been demonstrated in a laboratory setting.\cite{fdmspica2016}
Thus using established approaches, the multiplexing unit must be replicated $\sim$~10,000 times to meet the requirements of CMB-S4, illustrating the importance of a high-multiplexing factor readout scheme.

An alternative to TES bolometers is microwave kinetic inductance detectors (MKIDs) \cite{day2003}, which achieve high multiplexing factors by way of frequency division multiplexing many micro-resonators at gigahertz frequencies.
The microwave SQUID multiplexer (\umux)\cite{irwinlehnert2004, matesumux} borrows the multiplexing via microwave resonators feature of MKIDs, but couples to other types of sensors, such as TESs and metallic magnetic calorimeters.\cite{kempf2017}
As TES bolometers have been the workhorse sensor for CMB measurements in the last decade, TESs coupled to microwave SQUIDs in large numbers is an attractive technology for future CMB imagers.
The \umux\ is gaining maturity for both bolometric and calorimetric TES applications.\cite{stanchfield2016,mates2017} 

In this letter, we focus on the \umux\ for CMB polarization experiments.
Demonstrating key performance characteristics of the \umux\ on a small-scale is a necessary first step before developing a full-scale \umux\ capable of reading out 1000s of CMB polarimeters. 
We present results from a 64-channel multiplexer coupled to TES bolometers that have been optimized for CMB polarimetry. 
We show that the \umux\ is well-matched to the needs of ground-based CMB imagers in two important areas: noise performance and signal cross-talk.

\begin{figure}
    \centering
    \includegraphics[width=1.0\linewidth]{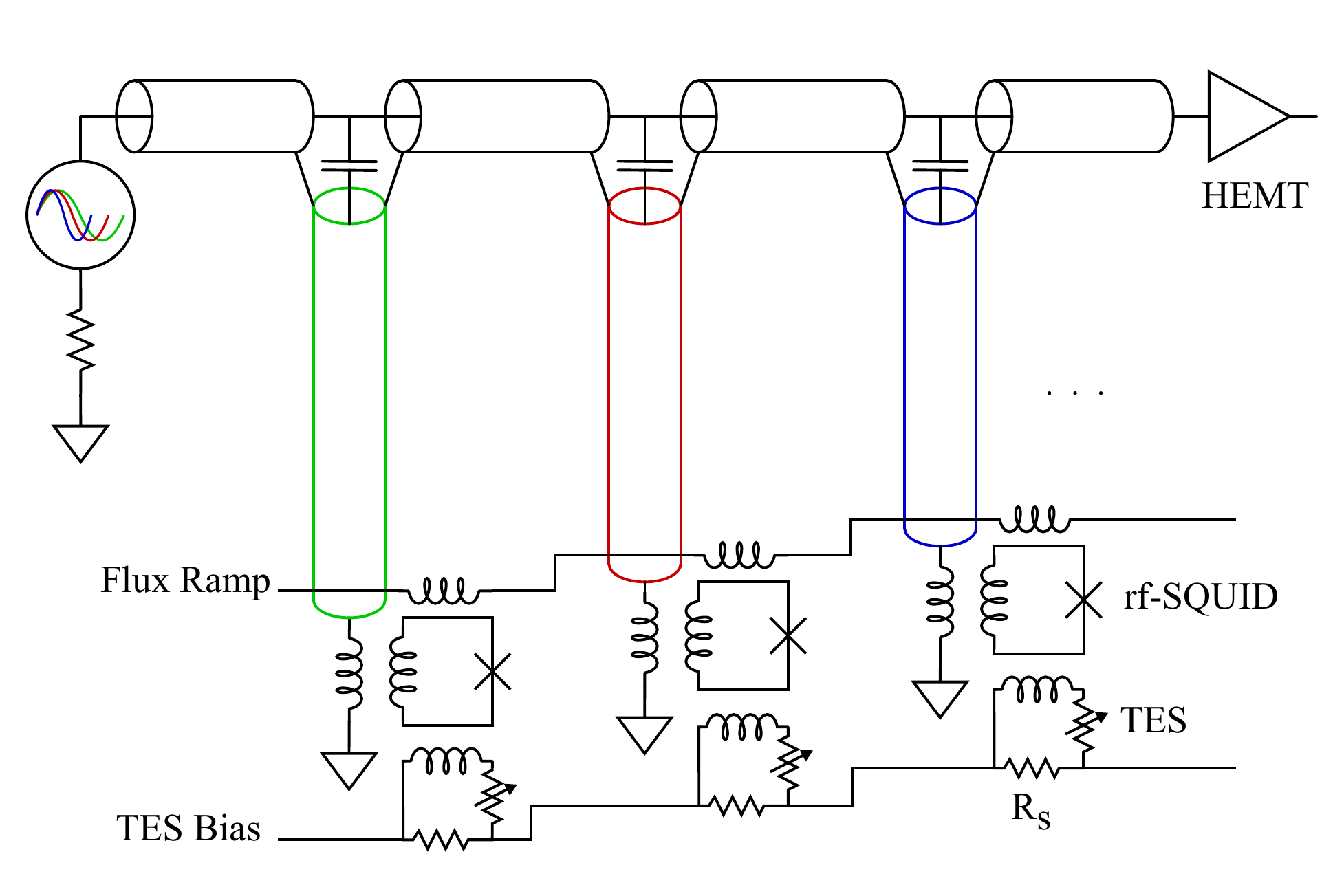}
    \caption{Conceptual schematic of a three-channel microwave SQUID multiplexer (\umux) for the readout of transition-edge-sensor (TES) bolometers. One multiplexer unit, which in principle reads out $>$1000 sensors, requires a pair of coaxial cables and two pairs of DC lines, which are used for detector bias and flux-ramp modulation.
    }
    \label{fig:umuxschem}
\end{figure}

The \umux\ readout is shown schematically in Fig.~\ref{fig:umuxschem}.
TES bolometers couple to rf-SQUIDs that inductively load high-$Q$ microwave resonators.
Signals sourced from each TES modulates the resonant frequency of a unique micro-resonator.  
All resonators couple weakly to a common microwave transmission line.  
Interrogating each resonator with a probe tone monitors the resonant frequency and therefore the TES current.  
Probe tones are amplified by a wide-bandwidth, low noise amplifier following the \umux\ circuit.
Flux-ramp-modulation avoids per channel feedback wires, encodes signal to the phase of the SQUID's periodic response, and requires a DC line common to all readout channels.\cite{mates2012flux}
Enabled by the high-$Q$ resonators at around gigahertz frequencies, multiplexing factors in excess of 1000 are achievable, and in practice are limited by resonator frequency spacing (constrained by signal bandwidth and allowable cross-talk) and the total bandwidth of the room temperature electronics.  
Further details on the \umux\ may be found in several previous publications.\cite{jabmates,noroozianumux,matesumux} 

\begin{figure}
    \centering
    \includegraphics[width=1.0\linewidth]{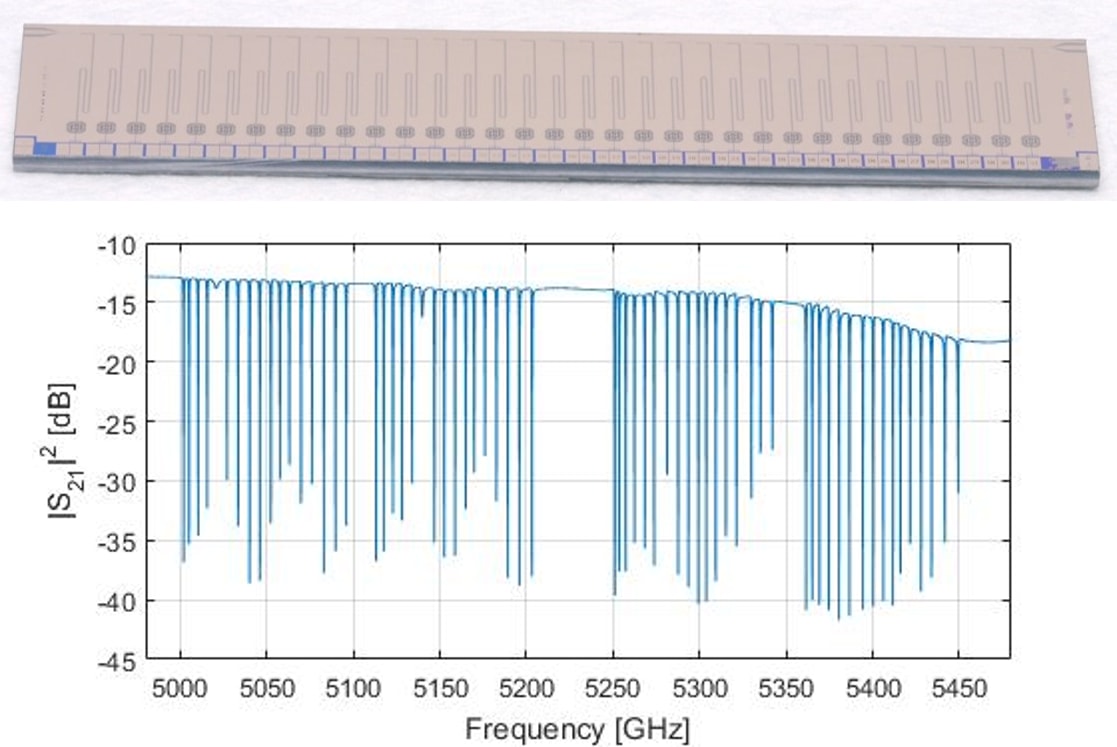}
    \caption{Above: A 32-channel \umux\ readout chip. Below: The $\mathrm{S}_{21}$ transmission of two distinct 32 channel \umux\ connected in series to produce a 64 channel multiplexer.
    }
    \label{fig:umuxs21}
\end{figure}

For this demonstration, we mount four detector chips, each containing six TES bolometers, into a sample box together with two 32-channel \umux\ die-wired in series, creating a 64-channel multiplexer.
The input coil that couples the TESs to the rf-SQUID has a mutual inductance of 230~pH.
The sample box is mounted to the 100~mK stage of a $^3$He-backed adiabatic demagnetization refrigerator (ADR). 
An image of a single \umux\ chip, as well as the transmission of the two \umux\ chips used in this letter, is shown in Fig. \ref{fig:umuxs21}.
The \umux\ channels are spaced 6~MHz apart with resonant frequencies between 5 and 6~GHz.  
Each resonator has a bandwidth of 300~kHz set by the strength of the capacitive coupling to the common feedline.
The flux ramp rate is 62.5~kHz, which in this implementation, sets the effective detector sampling rate to the same value.  
The output of the \umux\ is amplified by a High-electron-mobility transistor (HEMT) amplifier mounted at 4~K with a noise temperature $T_N \approx $~3~K.
To monitor current from each bolometer, we use the same software defined radio-based room temperature electronics as described in Mates et al.\cite{mates2017} 

The bolometer design is similar to the devices implemented in Advanced ACTPol arrays.\cite{ho2016}
Each bolometer contains a 25~$\mu$m~$\times$~200~$\mu$m, 350~nm thick Al-Mn TES which produces a normal resistance $R_n \approx 8.2~ m\Omega$ and superconducting transition temperature $T_c \approx 180$~mK. The TESs are thermally isolated from the substrate by four SiN legs of dimension 61~$\times$~20$~\times$~2~$\mu$m$^3$.
We determine the thermal transport properties by measuring the Joule power required to keep the bolometers in-transition as a function of the substrate temperature, $T_b$, and by fitting to
\begin{equation}
    \label{eq:nKT}
    P(T_b)=K(T^{n}-T_b^{n}).
\end{equation}
The conductance exponent $n$, prefactor $K$, and TES temperature $T$ are free parameters. 
When fitting Eq. \ref{eq:nKT} over $T_b$ = 50-170~mK in 15~mK steps, we find $n=3.6$, $K=8765~\mathrm{pW}/$K, $T=182.5$~mK, and the thermal conductance $G=nKT^{n-1}=~$390~$\mathrm{pW}/$K.  
All bolometers incorporate 0.5~$\Omega$ Au resistors, which are used to inject known signals for cross-talk measurements.  
The TESs are biased on a common line via two 32-channel shunt resistor chips with $R_s$~=~374~$\mu\Omega$, determined from a 4-point resistance measurement at 100~mK. 

CMB instruments are designed to maximize sensitivity, and therefore detectors are ideally limited only by fundamental noise sources.  
There are two fundamental noise sources to consider: photon noise and detector noise. 
The photon noise, noise equivalent power (NEP) for a single mode of a narrow-band source is \cite{zmuidzinas2003thermal}
\begin{equation}
    \label{eq:photonNoise}
    NEP_\gamma~=~\sqrt{2h\nu P_o(1+m\eta)},
\end{equation}
where $\nu$ is the source center frequency, $\eta$ is the optical efficiency of the system, $P_o$ is the radiative power absorbed in the detector from all sources, $m$ is the photon occupation number, and $h$ is Planck's constant. 
The fundamental noise term associated with the detector is thermal fluctuation noise arising from phonon fluctuations in the bolometer weak thermal link.  
To obtain an expression of phonon noise, we assume the radiative transfer model of Boyle and Rogers\cite{boyle1959performance} which we find to be a good match for our devices.\cite{duff2016advact,hubmayr2016spider}
When recast in terms of the variables defined in Eq.~\ref{eq:nKT}, we obtain the following expression for phonon NEP,
\begin{equation}
    \label{eq:gnoise}
    NEP_{G}=\sqrt{2k_{b}T Pn\frac{1+(T_{b}/T)^{n+1}}{1-(T_{b}/T)^{n}}}.
\end{equation}

The readout noise must be sub-dominant to the quadrature sum of $NEP_{\gamma}$ and $NEP_{G}$.  
Our readout system is sensitive to current sourced from the TES bolometers, and thus to compare measurement to these fundamental noise source terms, we must convert the $NEP$ terms to current noise by use of the device responsivity $\delta I / \delta P$.  
For a high loop gain TES bolometer ($\mathcal{L}~=P\alpha / GT>>1$, $\alpha$ is the logarithmic sensitivity of the transition with respect to the temperature), $\delta I / \delta P \approx 1 / V$, where $V$ is the voltage across the TES.  
For these devices $\mathcal{L}~\sim$~20, and thus the approximation is valid. 

To evaluate the noise performance of the multiplexer, we measure the current noise of the system in three states.  
In one condition, readout channels are not coupled to TES bolometers and represent the readout noise floor.  
In a second condition, we couple readout channels to TES bolometers that are unbiased and thus in their superconducting state.  
In the third condition, we voltage-bias the detectors into the transition between the superconducting and normal state.  
To mimic a photon load, we elevate $T_b$ from the nominal 100~mK to 150~mK, producing a roughly one-to-one ratio of photon load to Joule power ($P_o$~=~7.6~pW, $P(T_b~=~150~mK)~=~10.04$~pW) typical of CMB observations.
This action ensures that the device responsivity is representative of the value during true observations.
For each measurement state, we acquire 100-second detector timestreams, which are used to produce the noise spectra.      

\begin{figure*}
    \centering
    \includegraphics[width=.8\linewidth]{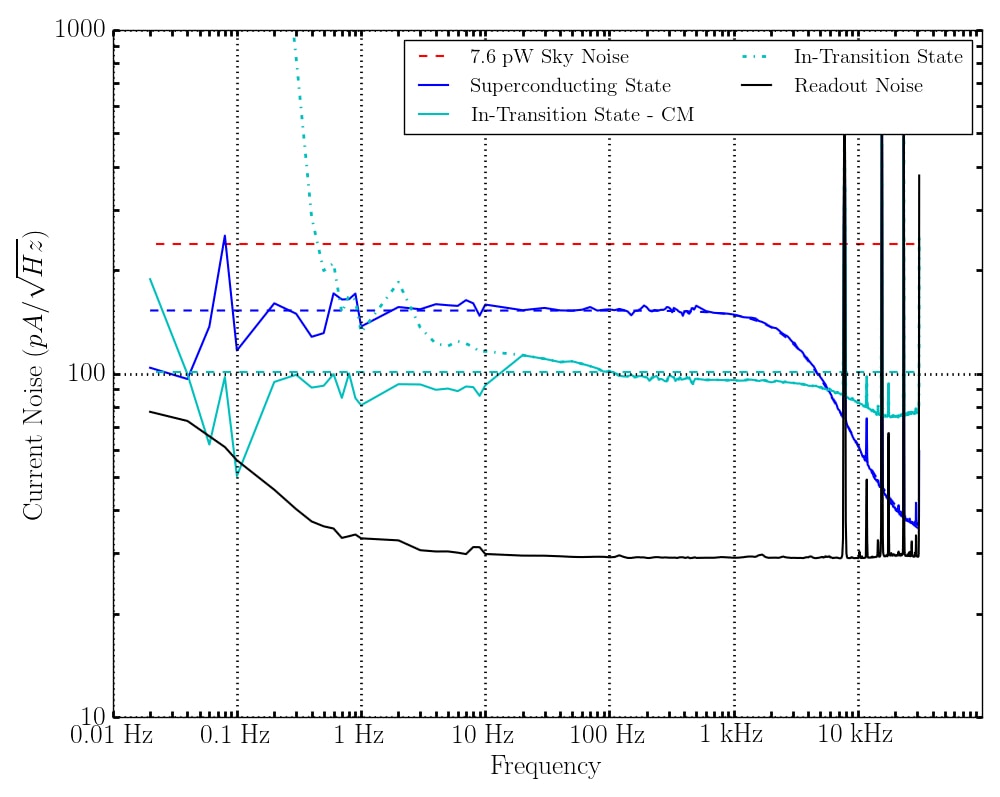}
    \caption{Cumulative set of \umux\ current noise measurements.  
    The average of 35 open input channels (black, solid line) demonstrates a 29~pA$/\sqrt{\mathrm{Hz}}$ readout noise floor.  
    A spectrum of a TES-coupled channel in the superconducting state (solid, blue) matches predictions (dashed, blue) described in the text.  
    Noise measured for a detector biased into the superconducting transition is shown in teal for raw data (dot-dashed curve) and post common-mode subtraction (solid), which removes temperature drifts.  
    The noise level matches expectations from fundamental thermal fluctuation noise in the bolometer (dashed teal line), and exhibits a white spectrum at 100~mHz.  
    For illustration purposes, we show that the expected level of photon noise from a 7.6~pW thermal source at 150~GHz is a factor of 8 above the readout noise floor. 
    Spikes near 10~kHz are due to the warm readout electronics and are well understood. 
    }
    \label{fig:noise}
\end{figure*}

Fig.~\ref{fig:noise} displays the noise results. 
The averaged noise of the 35 readout channels that are not connected to TES bolometers exhibit a white readout noise floor of $I_{RN}~=~29~\pm~$2.6~pA$/\sqrt{\mathrm{Hz}}$.  
This noise level is known to be a function of microwave tone power.  
For example, $I_{RN}$~=~17~pA$/\sqrt{\mathrm{Hz}}$ has previously been demonstrated with $\sim$-70~dBm tone power.\cite{bennett2015}
However, the noise level shown here is sufficient for our purposes, as we will show.
The increased noise at f$<$2~Hz is due to room temperature electronics and is not fundamental to the device.

A measurement of one readout channel coupled to a TES in the superconducting state is shown in blue.  
In this state, noise from the TES shunt resistor $R_s$ contributes in addition to the readout noise. 
Explicitly, we expect
\begin{equation}
    \label{eq:shunt}
    I_{R}=\sqrt{\frac{4k_{b}T/R_s}{1+(\omega\tau)^{2}}+I_{RN}^{2}},
\end{equation}
where $\tau=L/R_s$ and $L$ is the inductance in the TES bias loop.
The blue dashed line is the predicted noise spectrum for the superconducting state, using Eq.~\ref{eq:shunt} and assuming $L$~=~95~nH.  
The data are in excellent agreement with the model, which shows that the sensor bias circuit is well-understood and that no unaccounted noise sources exist.  

Lastly, we measure the noise of the system when operating the bolometer in-transition at 0.65$R_n$.
The averaged measured white noise level between 5-50~Hz is 98.1~pA/$\sqrt{\mathrm{Hz}}$, consistent with expectations from the quadrature sum of readout noise and bolometer thermal fluctuation noise (using Eq.~\ref{eq:gnoise} with $P$=10.04~pW, $V$~=~233~nV, and the previously stated thermal conductance parameters, the Noise Equivalent Current, $NEI_{G}$~=~96~pA/$\sqrt{\mathrm{Hz}}$).  
As such the system noise level is 3\% higher than the bolometer thermal fluctuation noise level.  
When coupled to the sky, photon noise dominates the noise budget for an optimized bolometer.  
In the case of $P_o$=7.6~pW at $\nu$~=~150~GHz ($NEP_{\alpha}$ shown in Fig.~\ref{fig:noise}, red dashed line), the \umux\ produces a system noise level that is only 1\% greater than the level produced from the two fundamental noise sources.

After allowing for common mode subtraction, the measured low frequency noise performance of this readout system compares favorably with the anticipated needs of near-term CMB polarization experiments.  
The common mode subtraction uses a standard singular value decomposition routine that removes the three lowest order modes common to all detector timestreams.  
These modes are sourced by bath temperature drifts and pulse-tube mechanical vibrations.   
The detector noise as read out by the \umux\ is white at frequencies above 100~mHz.
Stability on long-time scales enables access to larger patches of the sky, which is required for some of the CMB inflationary science goals.  
For example the inflationary recombination bump peaks at $\sim~2^\circ$ angular scales.  
The stability requirement is more stringent for a large-aperture telescope, which produces a small beam and has limited scan speed due to the telescope mass.    
Nevertheless, the Atacama Cosmology Telescope (ACT) scan strategy \cite{thornton2016} (1.5~deg/s azimuth scans at a mean elevation of 50$^\circ$), places the peak of the recombination bump at audio frequencies a factor of 7 above the 1/$f$ knee presented in Fig.~\ref{fig:noise}.  


Channel-to-channel crosstalk is an important systematic effect that ultimately limits the measurement sensitivity to cosmological parameters.  
Recent work using time-division multiplexing shows that 0.3\% cross-talk in pair-differenced, polarization-sensitive detectors produces a false detection of the tensor-to-scalar ratio ($r$)-- the determination of which is a major goal of the CMB field--to $\sim$~20 times below the current best upper limits.\cite{ade2015bicep2} 
Since the goal of future CMB instruments is to constrain $r$ to $\lesssim$~10$^{-3}$, signal crosstalk due to the readout must be $<$0.3\% at a minimum, and likely lower than this value.

An expected source of channel crosstalk in the \umux\ is due to the overlapping tails of adjacent resonator Lorentzian profiles. 
When using flux-ramp demodulation, this crosstalk is proportional to $(2\pi\times16n^{2})^{-1}$, where $n$ is the resonator spacing in number of resonator bandwidths.\cite{jabmates} 
For our current spacing of $n=$~20, this amounts to a crosstalk level of 0.025$\%$ for nearest frequency neighbor channels, and is negligible for channels separated further away in frequency space. 

\begin{figure}
    \centering
    \includegraphics[width=1.0\linewidth]{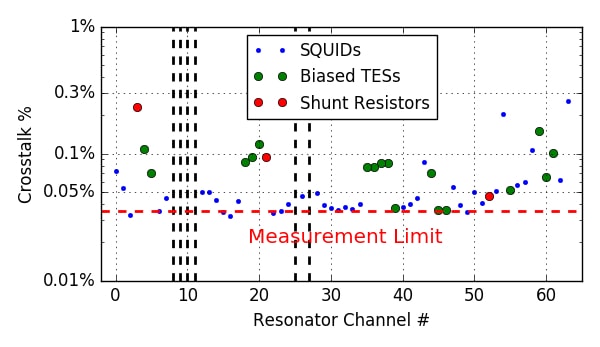}
    \caption{\umux\ channel crosstalk plotted versus resonator number (ascending resonant frequencies).  Vertical dashed lines represent the six perpetrator channels.  Green and blue circles correspond to TES-coupled and uncoupled readout channels, respectively.  The results show crosstalk is universally $<$0.3\%, and that 90\% of the channels are at or near the 0.03\% measurement floor, denoted by the horizontal dashed red line.}
    \label{fig:crosstalk}
\end{figure}

To quantify the level of crosstalk, we apply a rectified 50~Hz sine wave to the bolometer heater line, which is common to all six bolometers on a chip.  
This heater pulse dissipates a small signal in each of the six perpetrator bolometers, which leaves 18 bolometer victim channels and 40 dark SQUID victim channels.  
The level of crosstalk is determined by comparing the amplitude of the 50~Hz power spectral bin in the victim channels to the maximum of the six perpetrator channels. This conservative crosstalk estimate overlooks the contributions by the five lesser perpetrator channels.
Fig.~\ref{fig:crosstalk} plots the measured crosstalk as a function of resonator number, ordered in frequency ascension. 
Crosstalk is universally below 0.3\%, which is less than the level of cross-talk from multiplexed readout systems in deployed CMB imagers.
Furthermore, $\sim$90\% of the channels have a level $<$0.05\%.
Nearest neighbor frequency channels to perpetrators show cross-talk at the measurement sensitivity level of $\sim$0.03\%, which is consistent with the stated Lorentzian prediction.
Victim channels connected to TES bolometers have on average a 70\% elevated cross-talk relative to SQUID victim channels.  
This is likely due to a combination of a larger pickup loop in the TES channels from necessary wiring between sensors and the readout, and thermal crosstalk which only affects TES channels as SQUIDs are largely insensitive to substrate temperature fluctuations.
A detailed explanation of the cross-talk matrix is the subject of future research.

In summary, our 64-channel demonstration shows that \umux\ achieves the necessary low noise and crosstalk for CMB polarization applications.  
White noise at $\sim$~100~mHz is particularly encouraging, because stability on long time-scales is an important metric for almost all bolometric applications.  
Crosstalk has been shown to be at or better than the level achieved with existing multiplexers.
This work motivates demonstration on larger arrays.  
The current multiplexer has a multiplexing density of 167 channels/GHz.  
In future work, we aim to realize 500 channels/GHz, which when coupled to a room temperature electronics system with 4~GHz of bandwidth (several platforms are in development) produces a modular 2000 channel multiplexing unit.
A scalable, high multiplexing factor readout system will enable future arrays of CMB polarimeters.





\begin{acknowledgments}
We acknowledge the support of the NIST Innovations in Measurement Science and NASA APRA programs.
\end{acknowledgments}

\end{document}